\documentclass[aps,prb,reprint,groupedaddress,showpacs]{revtex4-1}

\usepackage{graphicx}
\usepackage{dcolumn}
\usepackage{bm}

\begin{document}


\title{Fluctuation Theorem and Microreversibility in a Quantum Coherent Conductor}

\author{Shuji Nakamura$^{1}$, Yoshiaki Yamauchi$^{1}$,
Masayuki Hashisaka$^1$, Kensaku Chida$^1$, Kensuke Kobayashi$^1$}
\email{kensuke@scl.kyoto-u.ac.jp} 
\author{Teruo Ono$^1$ Renaud Leturcq$^2$, Klaus Ensslin$^3$, Keiji Saito$^4$,
Yasuhiro Utsumi$^5$, Arthur C. Gossard$^6$}

\affiliation{$^1$Institute for Chemical Research, Kyoto University, Uji,
Kyoto 611-0011, Japan.}

\affiliation{$^2$Institute of Electronics, Microelectronics and
Nanotechnology, CNRS - UMR 8520, Department ISEN, Avenue Poincar{\'e},
F-59652 Villeneuve d'Ascq, France.}

\affiliation{$^3$Solid State
Physics Laboratory, ETH Z\"{u}rich, CH-8093 Z\"{u}rich, Switzerland.}

\affiliation{$^4$Graduate School of Science, University of Tokyo, Tokyo
113-0033, Japan.}

\affiliation{$^5$Department of Physics Engineering, Mie University, Mi-e
514-8507, Japan.}

\affiliation{$^6$Materials Department, University of California, Santa
Barbara, California 93106, USA,}

\begin{abstract}
Mesoscopic systems provide us a unique experimental stage to address
nonequilibrium quantum statistical physics. By using a simple tunneling
model, we describe the electron exchange process via a quantum coherent
conductor between two reservoirs, which yields the fluctuation theorem
(FT) in mesoscopic transport. We experimentally show that such a
treatment is semiquantitatively validated in the current and noise
measurement in an Aharonov-Bohm ring. The experimental proof of the
microreversibility assumed in the derivation of FT is presented.
\end{abstract}

\date{\today}



\maketitle 

\section{Introduction}
Since the 1980's mesoscopic conductors have been serving as an ideal
stage to investigate the quantum scattering problem both theoretically
and experimentally, because the quantum transport through a single site
can be precisely probed in electronic measurement~\cite{DattaETMS}. The
Landauer-B{\"u}ttiker formalism embodies this advantage of mesoscopic
physics, as was successfully applied to the Aharonov-Bohm ring, the
quantum point contact, and the quantum dot, through which the mesoscopic
physics has been established [see Fig.~1(a)]. Not only the current
averaged over for a certain time ($\langle I \rangle$), but also the
current fluctuation $\langle(\delta I)^2 \rangle$ due to the partition
process is treated in the same
framework~\cite{ButtikerPRB1992,MartinPRB1992,BlanterPR2000}.  Actually,
the quantum shot noise measurement was successfully demonstrated, for
example, to provide the direct proof of the fractional
charge~\cite{de-PicciottoNature1997,SaminadayarPRL1997} and the Cooper
pair~\cite{JehlNature2000} by looking at how carriers are scattered at a
mesoscopic conductor.

These days the mesoscopic transport is invoking much interest from
another point of view. As the electron transport can be viewed as the
electron exchange process between the reservoirs via the conductor as
shown in Fig.~1(b), it serves as a well-defined test stage for
nonequilibrium quantum statistical physics~\cite{EspositoRMP2009}.  The
unique advantage of this approach lies in that the degree of
nonequilibrium can be finely tuned by the bias voltage applied to the
conductor. In addition, many events, namely numerous electron exchange
processes, can be monitored, which enables us to perform precise
measurement.

\begin{figure}[tbp]
\center
\includegraphics[width=.99\linewidth]{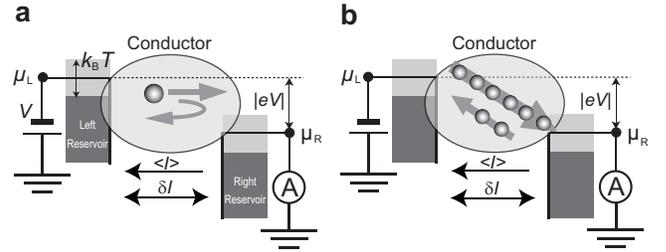}
\caption{(a) Schematic picture of the mesoscopic conductor coupled to the two
reservoirs in the nonequilibrium regime. $\mu_L$ and $\mu_R$ are the
chemical potentials of the left and the right reservoirs, respectively.
Mesoscopic transport based on the Landauer-B{\"u}ttiker picture is
schematically shown. When a conductor is biased, electrons injected from
one of the reservoirs are either transmitted or reflected at the
conductor, which yields finite current fluctuation [$\langle(\delta
I)^2\rangle\neq 0$].  (b) The transport can be also viewed as the
electron exchange process between the two reservoirs.}
\end{figure}

To quantitatively address the above topic, the fluctuation theorem
(FT)~\cite{EvansPRL1993} is believed to play a central
role~\cite{EspositoRMP2009,TobiskaPRB2005,AndrieuxJCP2004,AndrieuxNJP2009,SaitoPRB2008,ForsterPRL2008,Forster_arXiv,UtsumiPRB2010,CampisiPRL2010,AltlandPRB2010}. Based
on microscopic reversibility (``microreversibility'' or detailed
balance), this relation exactly links the probabilities of the
production and consumption of the entropy in a given system that is
coupled to the reservoir. FT corresponds to a microscopic expansion of
the macroscopic second law of thermodynamics and is proven to yield the
linear-response theory~\cite{GallavottiPRL1996}, and the Onsager-Casimir
relations~\cite{SaitoPRB2008}.  FT was experimentally proved to be valid
in classical systems such as a colloidal particle in
fluid~\cite{WangPRL2002} and a resistor~\cite{GarnierPRE2005}. Although
it was extended to the quantum regime~\cite{KurchanArxiv}, an
experimental check in this regime was still lacking.  More recently, FT
was theoretically addressed in the mesoscopic
transport~\cite{TobiskaPRB2005,AndrieuxJCP2004,AndrieuxNJP2009,UtsumiPRB2010,CampisiPRL2010,AltlandPRB2010}
even in the presence of the magnetic
field~\cite{SaitoPRB2008,ForsterPRL2008,Forster_arXiv} and was indeed
shown to be relevant in the analysis~\cite{UtsumiPRB2010} of the
electron counting
experiments~\cite{GustavssonPRL2006,FujisawaScience2006}. While the
incoherent tunneling events across the quantum dot(s) were investigated
in the above experiments, the validity of FT in the quantum coherent
regime was left to be addressed.

Recently, we experimentally showed the presence of nontrivial relations
between the nonlinear response and the nonequilibrium fluctuation in the
coherent transport of an Aharonov-Bohm (AB)
ring~\cite{NakamuraPRL2010}. When the current $I$ and the current
fluctuation (current noise power spectral density) $S$ are expanded in
the Taylor series as a polynomial of the bias voltage $V$,
\begin{equation}
I(V, B) = G_1(B) V + \frac{1}{2!} G_2(B) V^2 + \frac{1}{3!} G_3(B) V^3 + \cdots,
\label{EqPolyI}
\end{equation}
and
\begin{equation}
S(V, B) = S_0(B) + S_1(B) V + \frac{1}{2!} S_2(B) V^2 + \cdots,
\label{EqPolyS}
\end{equation}
we showed that there are proportional relations of $S_1^S \propto G_2^S$
and $S_1^A \propto G_2^A$. Here, the coefficients that are symmetrized
($S$) or antisymmetrized ($A$) with respect to the magnetic field
reversal are defined as 
\begin{equation}
G_2^{S, A} (B)\equiv G_2 (B) \pm G_2 (-B),
\end{equation}
 and
\begin{equation}
S_1^{S, A} (B)\equiv S_1 (B) \pm S_1 (-B)
\end{equation}
(take $+$ and $-$ for $S$ and $A$, respectively). This result is beyond
the consequence of the fluctuation-dissipation theorem $S_0(B) = 4 k_B T
G_1 (B)$.  Our observation semiquantitatively agrees with the
theoretical prediction on the basis of FT~\cite{SaitoPRB2008} and
provides an evidence of FT in the nonequilibrium quantum regime.

In this paper we expand the above work to further support our previous
report.  In Sec.~II, based on a simple tunneling model, we derive FT in
an applicable form to simple mesoscopic conductors. In Sec.~III, we
discuss the breakdown of the Onsager-Casimir reciprocity in the
nonequilibrium regime in the presence of the magnetic field. Then, as a
fundamental aspect of FT in mesoscopic transport, we show that the
validity of the microreversibility can be directly addressed in a
quantum regime.

\section{Fluctuation Theorem in a Mesoscopic System}
\subsection{Zero magnetic field case}
We explain FT by using the simplest setup and deduce the aforementioned
nonequilibrium fluctuation relations.  We consider a mesoscopic
conductor, say a quantum point contact, where the two quantum wires are
coupled by tunneling.  While more systematic and general derivation for
these relations is performed by using a cumulant generating
function~\cite{AndrieuxNJP2009,UtsumiPRB2010,CampisiPRL2010,AltlandPRB2010,SaitoPRB2008,ForsterPRL2008,Forster_arXiv},
the present simplest model is sufficiently instructive to treat here.
We assume that no energy relaxation takes place inside the conductor,
which is fulfilled in many mesoscopic devices smaller than the energy
relaxation length such as a quantum dot, chaotic cavity, ring, and so
on. First we treat the zero-magnetic field case to show $S_1 = 2 k_B T
G_2$ in Eqs. (\ref{EqPolyI}) and (\ref{EqPolyS}). The relations between
the coefficients in the current and the current noise are schematically
shown in Fig.~2.

\begin{figure}[tbp]
\center
\includegraphics[width=.99\linewidth]{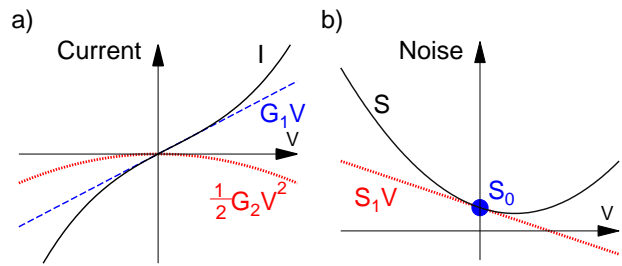}
\caption{ (color online) (a) Current-voltage characteristic as a
function of the bias voltage $V$. While Ohm's law holds around $V = 0$,
the current $I$ is not linear at large $V$ and the $I$-$V$ characteristics
can be decomposed into a polynomial of $V$ with coefficients $G_1$,
$G_2$, $G_3 \dots$ as in Eq.~(\ref{EqPolyI}).  In this study we address
$G_2$.  This schematic graph shows the total current $I$, the $G_1V$
contribution, and the $1/2! G_2 V^2$ contribution, in the solid, dashed,
and dotted curves, respectively.  The case with a negative $G_2$ is
shown.  (b) Similarly, the current noise spectral density $S$ (shown in
the solid curve) can be expressed in a polynomial form of $V$ as in
Eq.~(\ref{EqPolyS}). The Johnson-Nyquist relation tells that $S(V = 0) =
S_0 = 4k_BTG_1$.  This schematic graph shows the case with a negative
$S_1$ in the dashed curve. Here we address the coefficient of the term
linear in $V$ ($S_1$) and the relation between $S_1$ and $G_2$. }
\end{figure}

The present system is described by the following Hamiltonian:
\begin{eqnarray}
H &=& H_L + H_R + H_{LR},
\end{eqnarray}
where $H_L$ and $H_R$ are the Hamiltonian of the left and right quantum
wires and $H_{LR}$ is the tunneling part between them.  The initial
density matrix is decoupled into the equilibrium states of each wire,
where the left and right wires are assumed to have equal temperature
$1/\beta = k_BT$ and have chemical potentials $\mu_L$ and $\mu_R$,
respectively. Then the whole density matrix is described by
\begin{eqnarray}
\hat{\rho}_{\rm initial} &=& \sum_{n_L , n_R }  \rho_{n_L, n_R }
|n_L , n_R\rangle \langle n_L , n_R |  , \\ 
\rho_{n_L, n_R } &=& 
{e^{-\beta [E_{n_L} - \mu_{L} \, n_{L } ] } \over Z_{L}}
{e^{-\beta [E_{n_R} - \mu_{R} \, n_{R } ] } \over Z_{R}},
\end{eqnarray}
where $Z_L$ and $Z_R$ are the normalization factors, and $|n_L , n_R
\rangle$ defines the state that $n_L$ and $n_R$ electrons are present
inside the left and right wires with the eigenenergies $E_{n_L}$ and
$E_{n_R}$ of $H_L$ and $H_R$, respectively.  

The probability of finding the state $|n_L ' , n_R '\rangle $ after a
certain time $\tau$ starting from the initial state $|n_L , n_R \rangle$
is expressed as
\begin{eqnarray*}
P_{(n_L , n_R )\to (n_L' , n_R')} 
\!=\! |\langle n_L' , n_R' | e^{{-i\tau \over \hbar}H } | n_L , n_R \rangle |^2 
\rho_{n_L , n_R} . ~
\end{eqnarray*}
The microreversibility or the time reversal symmetry is given by
\begin{eqnarray*}
|\langle n_L' , n_R' | e^{{-i\tau \over \hbar } H } | n_L , n_R \rangle |^2  \!=\!
|\langle n_L , n_R | e^{{-i\tau \over \hbar }  H } | n_L', n_R'\rangle |^2  .~
\end{eqnarray*}
Here for simplicity we assume that $| n_L , n_R \rangle$ and $| n_L' ,
n_R' \rangle$ has the time reversal symmetry as the electron numbers are
the good quantum number, while in the general
treatment~\cite{SaitoPRB2008} this assumption is not necessary.  

As the electron number conservation $n_{L}- n_L' = -(n_R - n_R') $ and
the energy conservation are satisfied at very large $\tau$, $E_{n_L'}
-E_{n_L} \approx -(E_{n_{R}' } - E_{n_R } ) $. Using the
microreversibility and the conservation laws, we find the relation
\begin{eqnarray*}
P_{(n_L ,n_R)\to (n_L' , n_R') } 
&=& P_{(n_L' ,n_R')\to (n_L , n_R) }  e^{A (n_L - n_{L}')}, \label{dft}
\end{eqnarray*}
where $A$ is an affinity $A=\beta (\mu_L - \mu_R)$.  The probability
that the number of the transmitted electron is $Q$, is defined as $P(Q)
= \sum_{n_L , n_R , n_L' , n_R'}P_{(n_L ,n_R)\to (n_L' , n_R') } \delta
(Q - (n_L - n_{L}'))$.  Therefore, FT is obtained as the direct
consequence of the microreversibility
\begin{eqnarray}
P(Q) &=& P(-Q) e^{AQ} \label{ft}.
\end{eqnarray} 
This microreversibility ensures the following sum rule, which is called
``global detailed balance'' in Ref.~[\onlinecite{ForsterPRL2008}],
\begin{eqnarray}
\langle e^{AQ} \rangle = 1,
\label{EqGlobal}
\end{eqnarray} 
since $1 = \sum_Q P(Q) = \sum_Q P(-Q) e^{AQ} = \langle e^{AQ} \rangle$.
Here, $\langle\cdots\rangle$ denotes the expectation
$\langle\cdots\rangle \equiv \sum_Q \cdots P(Q)$.

Now let us discuss the higher order correlations between the current and
its noise power, which are the central topic in the present paper.  With
FT (\ref{ft}), we find the following identity
\begin{eqnarray}
\langle Q \rangle &=& \sum_Q Q P(Q) =
- \sum_Q Q P(Q) e^{-A Q} \nonumber \\
&=&
-\langle Q \rangle + A \langle Q^2 \rangle -{A^2\over 2! }\langle Q^3 \rangle +
\cdots . 
\end{eqnarray}
Furthermore, we note that $\langle Q^n \rangle$ can be expanded in the
Taylor series of $A$ with the coefficients $\langle Q^n \rangle_m$ ($n,
m$ integer)
\begin{eqnarray}
\langle Q^n \rangle &=& \langle Q^n \rangle_0 + A \langle Q^n \rangle_1
+ {A^2\over 2!} \langle Q^n \rangle_2 + \cdots .
\end{eqnarray}
Comparing order by order with respect to $A$, we find infinite number of
relationships among these coefficients, some of which are given as
\begin{eqnarray}
\langle Q^2 \rangle_0 &=& 2 \langle Q \rangle_1 , \label{1st}\\
 \langle Q^2 \rangle_1  &=& \langle Q \rangle_2  . \label{2nd}
\end{eqnarray}

Averaged current $I$ and current noise power $S$ are defined as
$I=\langle Q \rangle /\tau$ and $S = 2(\langle Q^2 \rangle - \langle Q
\rangle^2 )/\tau~$~\cite{NoteFactor2}.  The first relation (\ref{1st})
is equivalent to the fluctuation dissipation
relations~\cite{GallavottiPRL1996}
\begin{eqnarray}
S_0 = 4 k_{\rm B} T G_1,
\end{eqnarray}
and the second relation (\ref{2nd}) is to
\begin{eqnarray}
S_1 = 2 k_{\rm B} T G_2.
\label{S1G2}
\end{eqnarray}
This relation is beyond the fluctuation-dissipation relation and
directly links the nonlinearity and the nonequilibrium of the system.

\subsection{Finite magnetic field case}
At $B\neq 0$, the microreversibility requires that the probability
$P(Q,B)$ should satisfy~\cite{SaitoPRB2008}
\begin{eqnarray}
P(Q,B) = P(-Q, -B)\exp(AQ).
\label{PQB}
\end{eqnarray}
$P(Q,B)$ is now decomposed to the symmetric and antisymmetric parts
regarding the magnetic field reversal; $P_{\pm} (Q) \equiv P(Q,B) \pm
P(Q,-B)$, which fulfill
\begin{eqnarray}
P_{\pm}(Q) = \pm P_{\pm}(-Q)e^{AQ}.
\end{eqnarray}
Although the symmetric part $P_{+}(Q)$ produces the same fluctuation
relations as $P(Q)$ does, the antisymmetric probability gives rise to a
nontrivial result. By considering the antisymmetrized number of
charges exchanged between the reservoirs,
\begin{eqnarray}
\langle Q_{-} \rangle \equiv \sum_Q Q P_{-}(Q) = \sum_Q Q P_{-}(Q) e^{-AQ}
\end{eqnarray}
and defining $\langle Q_{-}^n \rangle_m$ with nonnegative integers $n$
and $m$ as the coefficients in the Taylor expansion of the above $\langle
Q_{-} \rangle$ with regard to $A$, we obtain
\begin{eqnarray}
\langle Q_{-}^3 \rangle_0 = 2\langle Q_{-}^2 \rangle_1.
\label{Asym1}
\end{eqnarray}
By noting the following relation, which is the consequence of the
normalization condition $\sum_Q P(Q)=1$,
\begin{eqnarray}
0 =  \sum_Q  P_{-}(Q) = - \sum_Q  P_{-}(Q) e^{-AQ},
\label{EqGlobalQ-}
\end{eqnarray}
we obtain
\begin{eqnarray}
3\langle Q_{-} \rangle_2 - 3\langle Q_{-}^2 \rangle_1 + \langle
 Q_{-}^3 \rangle_0 = 0. 
\label{Asym2}
\end{eqnarray}

The current that is antisymmetrized with regard to the $B$ reversal is
defined as $I(V, B)-I(V, -B) = \langle Q_{-} \rangle/\tau$ and the current
noise power is also defined in the same way, Eqs.~(\ref{Asym1}) and
(\ref{Asym2}) yields
\begin{eqnarray}
S_1^A =C_0^A/2k_BT 
\label{WMicroRev}
\end{eqnarray}
and
\begin{eqnarray}
S_1^A-2k_BTG_2^A = C_0^A/3k_BT,
\label{WOMicroRev}
\end{eqnarray}
respectively.  Here, $C_0^A$, which originates from the term $\langle
Q_{-}^3 \rangle_0$, is the antisymmetric part of the third cumulant at
equilibrium.  These two yield the antisymmetric relation expressed by
\begin{eqnarray}
S_1^A = 6k_BTG_2^A.
\end{eqnarray}

The above deduction totally relies on the microreversibility as is the
case in a systematic derivation~\cite{SaitoPRB2008}. Recently, however,
an interesting possibility of the broken microreversibility in
mesoscopic conductors is pointed
out~\cite{ForsterPRL2008,Forster_arXiv}.  It was discussed that, because
of the global detailed balance expressed by Eq.~(\ref{EqGlobal}), the
sum rule Eq.~(\ref{EqGlobalQ-}) and hence Eq.~(\ref{Asym2}) hold true
without microreversibility, even if we do not resort to the relation
$P_{-} (Q) = -P_{-} (-Q) \exp(AQ)$ (Eq.~(\ref{PQB})).  In this case,
Eq.~(\ref{Asym1}) and the resultant Eq.~(\ref{WMicroRev}) are no more
valid and only Eq.~(\ref{WOMicroRev}) is expected.  To address this
issue experimentally is the main motivation of the present paper.

The conventional current and shot noise formulas in the
Landauer-B{\"u}ttiker framework can be also expressed in the polynomial
form of $V$ \{see Eqs.~(39) and (61) in
Ref.~[\onlinecite{BlanterPR2000}]\}. By taking the energy-dependent
transmission into account, a relation similar to Eq.~(\ref{S1G2}) holds
true. However, this approach, which is based on the transmission defined
in the equilibrium, fails to explain the nonlinear conductance that is
not symmetric with respect to the magnetic field reversal. Indeed, it is
established theoretically and
experimentally~\cite{SanchezPRL2004,SpivakPRL2004,WeiPRL2005,AngersPRB2007,LeturcqPRL2006}
that due to electron-electron interactions induced in biased mesoscopic
conductors, the Onsager-Casimir reciprocal relations are broken, leading
to finite $G_2^A$. We will show the experimental data regarding this
below.

\section{Magnetic field asymmetry and microreversibility}

\subsection{Experiment}
We used an Aharonov-Bohm (AB) ring as a typical coherent conductor.
Figure~3(a) shows an atomic force microscope (AFM) image of the AB ring
fabricated by local oxidation using an AFM~\cite{HeldAPL1998} on a
GaAs/AlGaAs heterostructure two-dimensional electron gas (2DEG) (the
electron density $3.7 \times 10^{11}$ cm$^{-2}$, the mobility $2.7
\times 10^5$ cm$^2$/Vs and the electron mean free path $2.7$~$\mu$m).
The two-terminal current and noise measurement setup in a dilution
refrigerator is also shown in Fig.~3(a). The in-plane gates defined by
the oxide lines are grounded in the present measurement. The 2DEG has a
back gate to tune the electron density and the conductance of the AB
ring can be modulated by the back gate voltage $V_g$ and the magnetic
field $B$ by the AB effect. Figure~3(b) shows the image plot of the
conductance as a function of $V_g$ and $B$. The upper panel of Fig.~3(b)
presents the conductance at $V_g =-0.09$~V displaying clear AB
oscillations with an oscillation period being 25~mT in agreement with
the ring radius of 230~nm~\cite{LeturcqPRL2006,YamauchiPRB2009}. The
conductance of the ring ranges between 1.3 and 1.7 in units of $2e^2/h
\sim (12.9$~k$\Omega)^{-1}$ with typical visibility of $\sim 0.13$.  The
presence of electron interferences guarantees the coherent electron
transport in the device.

\begin{figure}[tbp]
\center
\includegraphics[width=.99\linewidth]{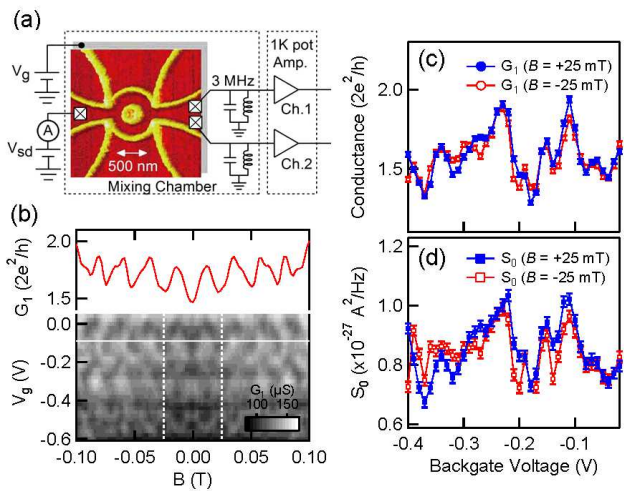}
\caption{ (color online) (a) Atomic force microscope (AFM) image of the
AB ring fabricated by local oxidation using an AFM~\cite{HeldAPL1998} on
a GaAs/AlGaAs 2DEG with the experimental setup for the two-terminal
current and noise measurements.  (b) Image plot of the conductance of
the ring as a function of $V_g$ and $B$ (in the lower panel). The upper
panel shows the magnetoconductance at $V_g=-0.09$~V as indicated by a
white line in the lower panel. (c) Conductance as a function of $V_g$ at
$B = \pm 25$~mT [indicated by dashed lines in the lower panel of
(b)]. (d) Equilibrium noise ($S_0$) as a function of $V_g$ at $B = \pm
25$~mT. }
\end{figure}

In addition to the dc measurement, we performed the noise measurement as
follows [also see Fig.~3(a)]. The voltage fluctuation across the sample
on the resonant circuit, whose resonant frequency is about 3.0~MHz with
the bandwidth of $\sim 140$~kHz, is extracted as an output signal of the
cryogenic
amplifier~\cite{de-PicciottoNature1997,DiCarloRSI2006,HashisakaPRB2008,NakamuraPRB2009,NakamuraPRL2010}. The
time-domain signal is then captured by the two-channel digitizer, and is
converted to spectral density data via FFT.  To increase the resolution
of the noise spectrum, we performed the cross-correlation technique by
using two sets of resonant circuit and amplifier. The sample was placed
in a dilution refrigerator whose base temperature is 45 mK and the
electron temperature in the equilibrium was 125 mK as deduced from the
thermal noise.  By numerically fitting the obtained resonant peak, the
current noise power spectral density $S$ is obtained as performed in
Ref.~\onlinecite{DiCarloRSI2006}.

In the analysis of the current $I$ and the current noise $S$ as
polynomials of $V$, the bias window was set to $|eV| \leq 50$~$\mu$eV,
which corresponds to $4.6 k_B T$ at $T=125$~mK.  In this bias range, the
Joule heating is expected to be negligible as seen in previous shot
noise measurements for mesoscopic
devices~\cite{HashisakaPRB2008,NakamuraPRB2009}. The coefficients in
Eqs.~(\ref{EqPolyI}) and ~(\ref{EqPolyS}) are deduced from the numerical
fitting to the obtained current and current noise.  The polynomial
fitting for $I$ and $S$ was performed by taking up to the fifth order of
$V$ for $I$ and up to the fourth order of $V$ for $S$ into account,
respectively. The analysis up to third or seventh order of $V$ for $I$
and second order of $V$ for $S$ yields results consistent with those
presented below. We note that the measurement was carefully performed at
several different $V_g$'s and $B$'s, and all the results are in a
quantitative agreement with each other within the experimental accuracy
of the present work.

\subsection{Results and Discussions}
Figure 3(c) shows the zero-bias conductance $G_1$ obtained at $B=25$~mT
and $B=-25$~mT at 125~mK as a function of the back gate voltage
$V_g$. Since $V_g$ modulates the electron density in the ring hence the
interference pattern, the conductance fluctuates as $V_g$ varies. As the
Onsager-Casmir reciprocity tells, $G_1$ behaves similarly at $B=25$ and
$-25$~mT as $V_g$ changes. The correlation factor ($CF$) between the
two, which is the covariance of the two divided by the product of their
standard deviations, is 0.91. Similarly, as shown in Fig.~3(d) the
gate-dependent thermal noises ($S_0$) at $B=25$~mT and $-25$~mT lap over
each other with $CF= 0.68$. Also we note that the proportionality
between $G_1$ and $S_0$ indicates that $S_0 = 4 k_BTG_1$ holds with an
electron temperature of $T= 125$~mK.  The coefficients of the first term
in Eqs.~(\ref{EqPolyI}) and ~(\ref{EqPolyS}) satisfy the Onsager-Casimir
reciprocity as a fundamental property in the equilibrium.

\begin{figure}[tbp]
\center
\includegraphics[width=.99\linewidth]{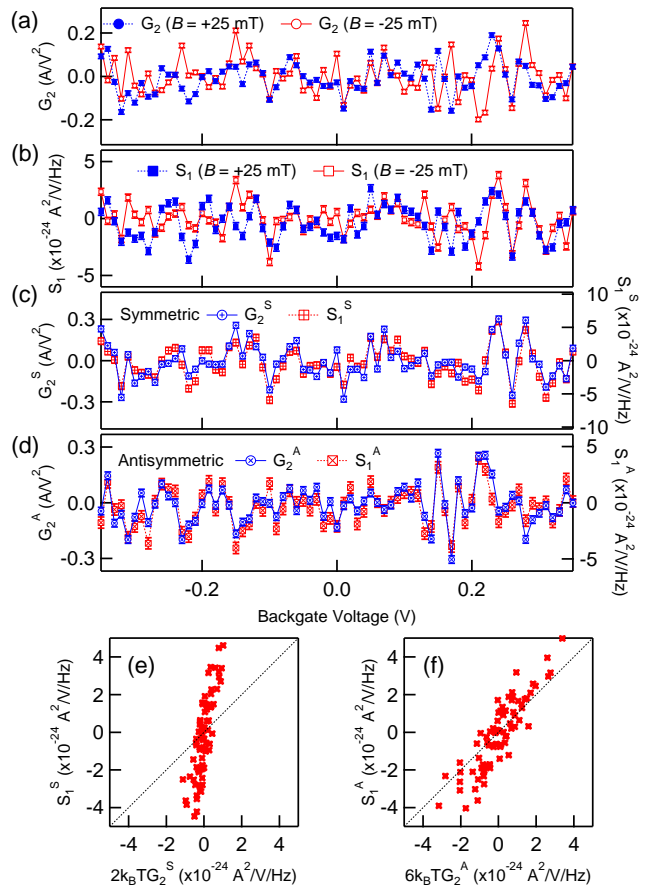}
\caption{ (color online) (a) $G_2$ obtained at $B=25$~mT and $-25$~mT is
shown as a function of $V_g$. (b) $S_1$ obtained at $B=25$~mT and
$-25$~mT is shown as a function of $V_g$. (c) The symmetrized components
$G_2^S$ and $S_1^S$ are shown in the left and right axes. (d) The
symmetrized components $G_2^S$ and $S_1^S$ are shown in the left and
right axes. (e) $S_1^S$ is plotted against $2 k_BT G_2^S$. The dotted
line is the prediction.  (f) $S_1^A$ is plotted against $6 k_BT
G_2^A$. The dotted line is the prediction. }
\end{figure}

Next we discuss the coefficients in the second term of
Eqs.~(\ref{EqPolyI}) and ~(\ref{EqPolyS}). Figures 4(a) and (b) show
$G_2$ and $S_1$ at $B=25$~mT and $-25$~mT, respectively. It is
remarkable that unlike the equilibrium property ($G_1$ and $S_0$), $G_2$
and $S_1$ are not symmetric with respect to the magnetic field
reversal. Indeed $CF$'s between the traces for the negative and positive
fields are as low as $0.20$ and $0.38$ for $G_2$ and $S_1$,
respectively. Regarding $G_2$, the presence of this asymmetry was
reported recently as the signature of the electron-electron correlation
effect induced in a biased mesoscopic
conductor~\cite{WeiPRL2005,AngersPRB2007,LeturcqPRL2006} The noise
measurement clearly tells that $S_1$ is also not symmetric with respect
to the magnetic field reversal.

Figures 4(c) and (d) show $G_2^S$ and $G_2^A$ in the left axis as a
function of $V_g$, respectively, where $S_1^S$ and $S_1^A$ are
superposed in the right axis.  Clearly, there appears strong correlation
between $G_2^S$ and $S_1^S$ and between $G_2^A$ and $S_1^A$ with $CF =
0.84$ and $0.85$, respectively.  As the theory predicts that $S_1^S =
2k_BTG_2^S$ and $S_1^A = 6k_BTG_2^A$, Fig.~4(e) and (f) shows the plots
to compare between the theory and the experiment. The dotted lines are
the prediction. As is consistent with the previous
report~\cite{NakamuraPRL2010}, the symmetric part deviates from the
theory while the antisymmetric part in Fig.~4(f) is in better agreement
with the theory than the symmetric one in Fig.~4(e). For the presented
data set, $S_1^S / 2k_BTG_2^S =6.00^{+0.94}_{-0.98}$ and $S_1^A
/6k_BTG_2^A = 1.61^{+0.22}_{-0.20}$, being statistically consistent with
the previous report~\cite{NakamuraPRL2010,PassingJCCCB1983}. The reason
for the observed considerable deviation from the theory in the symmetric
part is not yet clear. We note that in a double-quantum dot experiment
performed in the incoherent regime~\cite{FujisawaScience2006} similar
large discrepancy between the prediction based on FT was reported, where
the back action of the nonequilibrium quantum point contact attached to
the dots to detect their charge states explains the
observation~\cite{UtsumiPRB2010}. In the present case, as no such
detector is present, further effort is necessary to solve this problem.

Regarding the amplitude of $G_2^S$ and $G_2^A$, the experiment on the
nonlinear transport in the AB ring fabricated on the conventional 2DEG
was reported before~\cite{AngersPRB2007}. The radius of their ring is
about three times larger than ours. They measured the temperature
dependence of the amplitudes $G_2^S$ and $G_2^A$ and found that the
amplitudes rapidly decrease as temperature increases from 30~mK to 1~K.
Similar temperature dependence was observed in the present ring. At the
lowest temperature, the amplitude of $G_2^S$ and $G_2^A$ in the present
case is slightly larger but falls in the same range of their result.

\begin{figure}[htbp]
\center
\includegraphics[width=.99\linewidth]{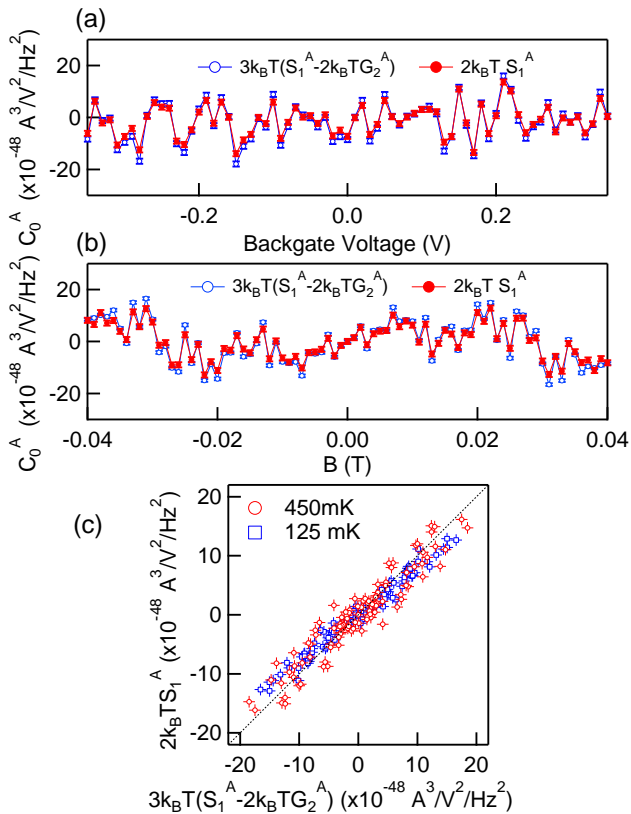}
\caption{ (color online) (a) Based on the data shown in Fig.~4(d) $3k_BT
(S_1^A-2k_BTG_2^A)$ and $2k_BT S_1^A$ are plotted as a function of
$V_g$. (b) Similar plot is shown for the data set in Fig. 3 in
Ref.~\onlinecite{NakamuraPRL2010}. (c) $3k_BT (S_1^A-2k_BTG_2^A)$ vs
$2k_BT S_1^A$ obtained at 125~mK and 450~mK.}
\end{figure}

Now let us discuss the microreversibility in the present system. In the
presence of the magnetic field, the possibility of the absence of the
microreversibility in the nonequilibrium was recently pointed
out~\cite{ForsterPRL2008}. While the antisymmetric relation is only
given by Eq.~(\ref{WOMicroRev}), the restriction of the
microreversibility simultaneously requires the relation of
Eq.~(\ref{WMicroRev}).  Thus we can basically obtain $C_0^A$ from the
experimental data in two ways; by calculating $C_0^A = 3k_BT
(S_1^A-2k_BTG_2^A)$ which holds true regardless of the
microreversibility condition and by calculating $C_0^A = 2k_BTS_1^A$
validated only with the microreversibility. Figure~5(a) shows the
obtained $C_0^A$ from the data set shown in Fig.~4. Clearly as a
function of $V_g$, $C_0^A$ calculated in two ways coincide each other.
Figure~5(b) shows the result for the data reported in Fig. 3 in
Ref.~\onlinecite{NakamuraPRL2010} where the field is swept with a fixed
$V_g$. In this case, too, $C_0^A$ obtained in two ways almost perfectly
equal each other.

In Fig.~5 (c), we plot $3k_BT (S_1^A-2k_BTG_2^A)$ vs. $2k_BTS_1^A$ at
125~mK and 450~mK. As we have seen that $S_1^A /6k_BTG_2^A$ deviates
from unity, the slope is slightly different from unity. However, within
the accuracy of the present experiment, we may claim that two values are
the same.  This tells us that in the present experiment the assumption
of the microreversibility is valid.

Finally, we note that the present demonstration gives a single example
of the validity of the microreversibility in the nonequilibrium quantum
regime in the presence of the magnetic field.  This fundamental topic
should be experimentally addressed in many systems such as electron
interferometers~\cite{ForsterPRL2008,Forster_arXiv,LimPRB2010}, the
quantum dot~\cite{SanchezPRB2009}, and the macroscopic inhomogeneous
system~\cite{NagaevPRL2010}.

\section{Conclusions}
We show that the fluctuation theorem is semiquantitatively valid in the
description of the quantum transport in mesoscopic systems. Unlike the
conventional scattering theory, this description gives a nontrivial
relation between the nonlinearity and the nonequilibrium in the
presence of the magnetic field. The direct test of the validity of the
microreversibility was also addressed.  Since the fluctuation theorem
does not directly give the physical interpretation of the current
through the device as the Landauer-B{\"u}ttiker formalism does, both
descriptions are complementary to each other. We believe that by
combining these two pictures, nonequilibrium properties in mesoscopic
systems in the presence of the interaction effect will be further
addressed.

We appreciate fruitful discussions from Markus B{\"u}ttiker, Masahito
Ueda, Takeo Kato, and Hisao Hayakawa. This work is partially supported
by KAKENHI, Yamada Science Foundation, SCAT, Matsuo Science Foundation,
Strategic International Cooperative Program the Japan Science and
Technology Agency (JST), and the German Science Foundation (DFG).


\begin{thebibliography}{99}

\bibitem{DattaETMS} See, e.g., S. Datta, {\it Electronic Transport in
Mesoscopic Systems} (Cambridge University Press, Cambridge, England,
1995); Y. Imry, {\it Introduction to Mesoscopic Physics} (Oxford UP, New
York, 1997).

\bibitem{ButtikerPRB1992} M. B{\"u}ttiker, Phys. Rev. B {\bf 46,} 12485 (1992).

\bibitem{MartinPRB1992} Th. Martin, R. Landauer, Phys. Rev. B {\bf 45,}
1742 (1992).

\bibitem{BlanterPR2000} Y. M. Blanter, and M. B{\"u}ttiker,
Phys. Rep.  {\bf 336,} 1 (2000).

\bibitem{de-PicciottoNature1997} R. de-Picciotto, M. Reznikov, M. Heiblum, V. Umansky,
G. Bunin, and D. Mahalu, Nature \textbf{389}, 162 (1997).

\bibitem{SaminadayarPRL1997} L. Saminadayar, D.C. Glattli, Y. Jin, B. Etienne,
Phys. Rev. Lett. \textbf{79}  2526 (1997).

\bibitem{JehlNature2000} X. Jehl, M. Sanquer, R. Calemczuk, and
D. Mailly, Nature \textbf{405,} 50 (2000).

\bibitem{EspositoRMP2009} M. Esposito, U. Harbola, and S. Mukamel,
Rev. Mod. Phys. {\bf 81,} 1665 (2009).

\bibitem{EvansPRL1993} D. J. Evans, E. G. D. Cohen, and G. P. Morriss,
Phys. Rev. Lett. {\bf 71,} 2401 (1993).

\bibitem{AndrieuxJCP2004} D. Andrieux, and P. Gaspard,
J. Chem. Phys. {\bf 121,} 6167 (2004); J. Stat. Mech.
P01011 (2006); J. Stat. Phys. {\bf 127,} 107 (2007).

\bibitem{TobiskaPRB2005} J. Tobiska, and Yu. V. Nazarov,
Phys. Rev. B  {\bf 72,} 235328 (2005).

\bibitem{AndrieuxNJP2009} D. Andrieux, P. Gaspard, T. Monnai, and
S. Tasaki, New J. Phys. {\bf 11,} 043014 (2009).

\bibitem{SaitoPRB2008} K. Saito, and Y. Utsumi,  Phys. Rev. B {\bf
78,} 115429 (2008); Y. Utsumi and K. Saito, {\it ibid.} {\bf 79,}
235311 (2009).

\bibitem{ForsterPRL2008} H. F{\"o}rster, and M. B{\"u}ttiker,
Phys. Rev. Lett. {\bf 101,} 136805 (2008).

\bibitem{Forster_arXiv} H.~F{\"o}rster, and M.~B{\"u}ttiker,
AIP Conference Proceedings {\bf 1129,} 20th International
Conference on Noise and Fluctuations, M.~Macucci and G.~Basso,
eds. (Melville, New York, 2009). p. 443.

\bibitem{UtsumiPRB2010} Y. Utsumi, D.~S. Golubev, M. Marthaler,
K. Saito, T. Fujisawa, and G. Sch{\"o}n, Phys. Rev. B {\bf 81,} 125331
(2010).

\bibitem{CampisiPRL2010} M. Campisi, P. Talkner, and P.
H{\"a}nggi, Phys. Rev. Lett. {\bf 105,} 140601 (2010).

\bibitem{AltlandPRB2010}
A. Altland, A. De Martino, R. Egger, and B. Narozhny
Phys. Rev. B {\bf 82,} 115323 (2010).

\bibitem{GallavottiPRL1996} G. Gallavotti, Phys. Rev. Lett. {\bf 77,} 4334
(1996).

\bibitem{WangPRL2002} G.~M. Wang, E.~M. Sevick, E. Mittag, D.~J. Searles, and
D.~J. Evans, Phys. Rev. Lett. {\bf 89,} 050601 (2002). 

\bibitem{GarnierPRE2005} N. Garnier and S. Ciliberto,  
Phys. Rev. E {\bf 71,} 060101(R) (2005).

\bibitem{KurchanArxiv} J. Kurchan, arXiv:cond-mat/0007360
(unpublished).

\bibitem{GustavssonPRL2006} S. Gustavsson, R. Leturcq, B. Simovic,
R. Schleser, T. Ihn, P. Studerus, K. Ensslin, D. C. Driscoll, and
A. C. Gossard, Phys. Rev. Lett. {\bf 96,} 076605 (2006).

\bibitem{FujisawaScience2006} T. Fujisawa, T. Hayashi, R. Tomita, and
Y. Hirayama, Science {\bf 312,} 1634 (2006).

\bibitem{NakamuraPRL2010} S. Nakamura, Y. Yamauchi, M. Hashisaka,
K. Chida, K. Kobayashi, T. Ono, R. Leturcq, K. Ensslin, K. Saito,
Y. Utsumi, and A. C. Gossard, Phys. Rev. Lett. {\bf 104,} 080602 (2010).

\bibitem{NoteFactor2} The factor 2 is artificially introduced so that
the resultant expression for $S_0$ and $G_1$ is consistent with the
classical expression of the Johnson-Nyquist relation.

\bibitem{SanchezPRL2004} D. S{\'a}nchez, and M. B{\"u}ttiker,
Phys. Rev. Lett. {\bf 93,} 106802 (2004); M. L. Polianski, and
M. B{\"u}ttiker, Phys. Rev. Lett. {\bf 96,} 156804 (2006); {\it ibid}
Phys. Rev. B {\bf 76,} 205308 (2007).

\bibitem{SpivakPRL2004} B. Spivak, and A. Zyuzin, Phys. Rev. Lett.
{\bf 93,} 226801 (2004).

\bibitem{WeiPRL2005} J. Wei, M. Shimogawa, Z. Wang, I. Radu,
R. Dormaier, D.H.  Cobden, Phys. Rev. Lett. {\bf 95,} 256601 (2005);
D. M. Zumb{\"u}hl, C. M. Marcus, M. P. Hanson, and A.C. Gossard, {\it
ibid.} {\bf 96,} 206802 (2006); C. A. Marlow, R. P. Taylor,
M. Fairbanks, I. Shorubalko, and H. Linke, {\it ibid.} {\bf 96,} 116801
(2006); B. Brandenstein-K{\"o}th, L. Worschech, and A. Forchel,
Appl. Phys. Lett. {\bf 95,} 062106 (2009).

\bibitem{AngersPRB2007}
L. Angers, E. Zakka-Bajjani, R. Deblock, S. Gueron, H. Bouchiat,  
A. Cavanna, U. Gennser, and M. Polianski, Phys. Rev. B {\bf 75,}
115309 (2007).

\bibitem{LeturcqPRL2006} R. Leturcq, D. S{\'a}nchez, G. G{\"o}tz,
T. Ihn, K. Ensslin, D. C.  Driscoll, and A. C. Gossard, Phys. Rev. Lett.
{\bf 96,} 126801 (2006); R. Leturcq, R. Bianchetti, G. G{\"o}tz, T. Ihn,
K. Ensslin, D.C. Driscoll, A.C. Gossard, Physica E {\bf 35,} 327-331
(2006).

\bibitem{YamauchiPRB2009} Y. Yamauchi, M. Hashisaka, S. Nakamura, K. Chida,
S. Kasai, T. Ono, R. Leturcq, K. Ensslin, D.
C. Driscoll, A. C. Gossard, and K. Kobayashi, Phys. Rev. B {\bf 79,}
161306(R) (2009).

\bibitem{HeldAPL1998} 
R. Held, T. Vancura, T. Heinzel, K. Ensslin, M. Holland, and
W. Wegscheider, Appl. Phys. Lett. {\bf 73,} 262 (1998).

\bibitem{DiCarloRSI2006} L. DiCarlo, Y. Zhang, D. T. McClure,
C. M. Marcus, L. N. Pfeiffer, and K. W. West, Rev. Sci. Instrum. {\bf
77,} 073906 (2006); M. Hashisaka, Y. Yamauchi, S. Nakamura, S. Kasai,
K. Kobayashi, and T. Ono, J. Phys. Conf. Series {\bf 109,} 012013
(2008).

\bibitem{HashisakaPRB2008} M. Hashisaka, Y. Yamauchi, S. Nakamura,
S. Kasai, T. Ono, and K. Kobayashi, Phys. Rev. B {\bf 78,} 241303(R)
(2008).

\bibitem{NakamuraPRB2009} S. Nakamura, M. Hashisaka, Y. Yamauchi,
S. Kasai, T. Ono, and K. Kobayashi, Phys. Rev. B {\bf 79,} 201308(R)
(2009).

\bibitem{PassingJCCCB1983} To estimate the error bars of the
coefficients between the two values, Passing-Bablok regression was
adopted since both $S_1$ ($S_1^S$ or $S_1^A$) and $G_2$ ($G_2^S$ or
$G_2^A$) have statistical uncertainties, where the conventional linear
regression is not justified to estimate the error bars. Here the error
bar indicates the 95\% confidence interval. H. Passing, and
W. A. Bablok, J. Clin. Chem. Clin. Biochem. {\bf 21,} 709 (1983).


\bibitem{LimPRB2010} J. S. Lim, D. S{\'a}nchez, and R. L{\'o}pez,
Phys. Rev. B {\bf 81,} 155323 (2010).

\bibitem{SanchezPRB2009} D. S{\'a}nchez, Phys. Rev. B 79, 045305 (2009).

\bibitem{NagaevPRL2010} K. E. Nagaev, O. S. Ayvazyan, N.Yu. Sergeeva,
and M. B{\"u}ttiker, Phys. Rev. Lett.  {\bf 105,} 146802 (2010).


\end{thebibliography}
\end{document}